\documentclass[12pt]{article}
\usepackage{times}
\usepackage{geometry}
\usepackage[utf8]{inputenc}
\usepackage{enumitem,amssymb}
\usepackage{ragged2e}
\usepackage{epsfig,colortbl}
\newlist{thematic}{itemize}{8}
\setlist[thematic]{label=$\square$}
\usepackage{pifont}
\newcommand{\OST}{{\em OST}}
\newcommand{\cmark}{\ding{51}}%
\newcommand{\done}{\rlap{$\square$}{\raisebox{2pt}{\large\hspace{1pt}\cmark}}%
\hspace{-2.5pt}}

\begin{document}
\raggedright
\huge
Astro2020 Science White Paper \linebreak

On the Origin of the Initial Mass Function \linebreak
\normalsize

\textbf{Thematic Areas:} \hspace*{60pt} $\square$ Planetary Systems \hspace*{10pt} $\done$ Star and Planet Formation \hspace*{20pt}\linebreak
$\square$ Formation and Evolution of Compact Objects \hspace*{31pt} $\square$ Cosmology and Fundamental Physics \linebreak
  $\done$  Stars and Stellar Evolution \hspace*{1pt} $\done$ Resolved Stellar Populations and their Environments \hspace*{40pt} \linebreak
  $\done$    Galaxy Evolution   \hspace*{45pt} $\square$             Multi-Messenger Astronomy and Astrophysics \hspace*{65pt} \linebreak
  
\textbf{Principal Authors:}

Name: Roberta Paladini	
 \linebreak						
Institution:  Caltech-IPAC
 \linebreak
Email: paladini@ipac.caltech.edu
 \linebreak
Phone:  + 1 626 395 1848
 \linebreak
 
 Name: Matthew Povich	
 \linebreak						
Institution:  California State Polytechnic University, Pomona
 \linebreak
Email: mspovich@cpp.edu
 \linebreak
Phone:  +1 909 869 3608
 \linebreak
 
\textbf{Co-authors:} (names and institutions)
  \linebreak
  Lee Armus (Caltech-IPAC), Cara Battersby (U. Connecticut), Bruce Elmegreen (IBM), Adam Ginsburg (NRAO), Doug Johnstone (NRC), David Leisawitz (NASA-GSFC), Peregrine McGehee (College of the Canyons), Sarah Sadavoy (CfA), Marta Sewilo (NASA-GSFC), Alessio Traficante (INAF), Martina Wiedner (Observatoire de Paris)\\

\hspace{0.5truecm}

\textbf{Abstract:}
It is usually assumed that the stellar initial mass function (IMF) takes a universal form and that there exists a direct mapping between this and the  distribution of natal core masses (the core mass function, CMF).  
The IMF and CMF have been best characterized in the Solar neighborhood. Beyond 500~pc from the Sun, in diverse environments where metallicity varies and massive star feedback may dominate, the IMF has been measured only incompletely and imprecisely, while the CMF has hardly been measured at all.
In order to establish if the IMF and CMF are indeed universal and related to each other, it is necessary to: 1) perform multi-wavelength large-scale imaging and spectroscopic surveys of different environments across the Galaxy; 2) require an angular resolution of ${<}0.1''$ in the optical/near-IR for stars and ${<}5''$ in the far-IR for cores; 3) achieve far-IR sensitivities to probe 0.1~M$_{\odot}$ cores at 2--3 kpc. 

\pagebreak
\section{Adopted forms of the IMF and the CMF}

The IMF is not an observable quantity, but rather an analytical description of the mass distribution among a newly-formed stellar population (Kroupa 2013).
Modern forms of the IMF adopt either a log-normal distribution at low masses and a power-law tail above 1~M$_{\odot}$ (Chabrier et al.\ 2003, 2005) or a continuous set of several ``broken'' power-laws (Kroupa 2001, 2013). Above 
$\sim$ 0.2 M$_{\odot}$, the Chabrier and Kroupa IMFs agree, and the integrated mass of a stellar population is the same using either IMF formalism (Chomiuk \& Povich 2011).  Below 0.2 M$_{\odot}$ the form of the IMF is still very uncertain and the subject of much debate.
    
~~~~~The CMF has a shape similar to Chabrier and Kroupa IMFs but is shifted towards larger masses by a factor ${\sim}3$, which has generally been interpreted as a core-to-star conversion efficiency of  ${\sim}30\%$ (see Fig.~2). 

~~~~~The similar shape of the IMF and CMF has led to believe that there is an intrinsic mapping between these two quantities. However, this one-to-one correspondence does not find much theoretical ground (see below).

\section{Are the the IMF and CMF Universal? }

Over the last decade, there has been growing evidence of a variable IMF, as opposed to the common assumption that the IMF of the Milky Way is universal (Kroupa 2002; Bastian et al.\ 2010, Fig.~1). These claims come from a wide variety of approaches, including stellar population analysis (e.g. van Dokkum $\&$ Conroy 2010; Ferreras et al.\ 2013), gravitational lensing (Treu et al.\ 2010), and dynamical models (Cappellari et al.\ 2012).
A notable Galactic example of exceptions to a universal IMF is the Taurus Molecular Cloud, which shows an excess of 0.6--0.8 M$_{\odot}$ stars. Other examples are the massive clusters Westerlund 1 (Lim et al.\ 2013), Quintuplet (Hussman et al.\ 2012), Arches (Hosek et al.\ 2019),  
and the young nuclear star clusters  (Lu et al.\ 2013), although these could depart from a {\em{standard IMF}} as a consequence of mass segregation.\\ 

~~~~~From the extragalactic point of view, since 2010 there has been a flurry of IMF studies 
focusing on early type elliptical galaxies. These studies have both found an over-abundance of high-mass stars (``top-heavy" IMF, e.g. 
Dav\'{e} et al.\ 2008), and an over-abundance of low-mass stars (``bottom-heavy" IMF, e.g. van Dokkum $\&$ Conroy 2010). 
In all cases, it is important to keep in mind that to determine the IMF of a stellar population, one has to go over a complicated process which consists of several steps: (1) measure the Luminosity Function (LF) of a complete sample of stars that lie in a defined volume; (2) convert the LF into a present day mass function (PDMF), using a mass-magnitude relationship; and (3) correct the PDMF for the star-formation history, stellar evolution, galactic structure, cluster dynamical evolution and binarity to obtain the individual-star IMF. Each of these steps is affected by potential biases and pitfalls that can lead to highly uncertain results.

\begin{figure}[h]
    \centering
   \includegraphics[width=0.55\textwidth]{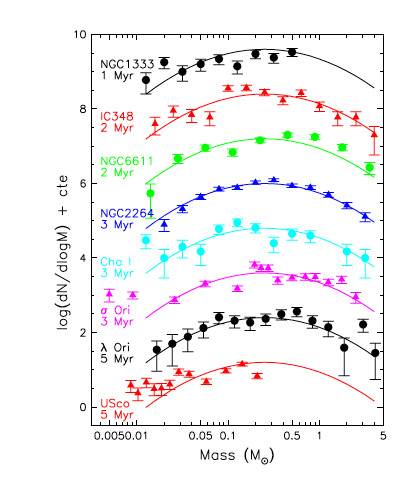}
    \caption{Recent IMF estimates for 8 star forming regions. The error bars represent the Poisson
error for each data point. The solid lines are the log-normal form proposed by Chabrier (2005) for the IMF, normalized to best follow the data. From Offner et al. (2014).}
    \label{fig:offner2014}
\end{figure}

~~~~~The most recent CMF determinations (Fig.~2) have been obtained with Herschel  (e.g., Andr\'{e} et al.\ 2010; Konyves et al.\ 2015, Olmi et al.\ 2018) and ALMA (Motte et al. 2018). The Herschel data support the conclusions of early studies performed in the $\rho$-Oph and Serpens molecular clouds (Motte et al.\ 1998; Johnstone et al.\ 2000; Testi $\&$ Sargent, 1998), which suggested that the CMF can be described, similarly to the IMF, by $dN \sim M_{\rm core}^{-1.5} dM$ below 0.5~M$_{\odot}$ and by $dN \sim M_{\rm core}^{-2 - 2.5} dM$ at higher core masses. However, the recent ALMA observations in the mini Galactic starburst W43 appear to show a departure from a standard IMF, with a much shallower CMF. \\

\begin{figure}[h]
    \centering
   \includegraphics[width=0.55\textwidth]{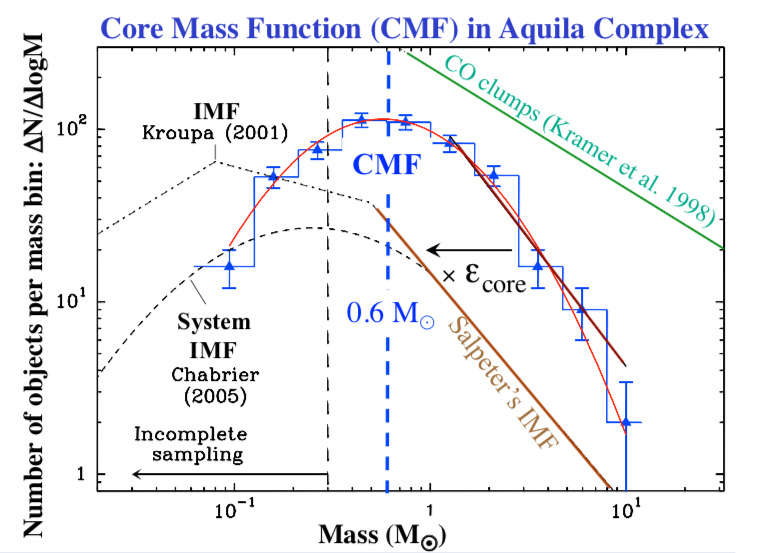}
    \caption{Core mass function (histogram with error bars) of the prestellar cores identified with Herschel in Aquila (Konyves et al.\ 2015; Andr\'{e} et al.\ 2010). The Kroupa and Chabrier IMF and the typical mass spectrum of CO clumps is shown for comparison.}
    \label{fig:andre2014}
\end{figure}

~~~~~From an analytical standpoint, Inutsuka (2001) and Hennebelle \& Chabrier (2008) applied the Press-Schechter formalism and Shadmehri \& Elmegreen (2011) used the ISM power spectrum with a density cutoff to obtain the clump mass function. A big uncertainty with this method, and with the conversion of a theoretical clump mass function into a stellar mass function, is the unknown multiplicity and mass function of stars inside each clump, for which there are few observations. Often clumps contain several stars and the one-to-one correspondence between clump mass and stellar mass is lost. Numerical simulations of star formation usually get the IMF in a more dynamical process involving long-term accretion into cores along filaments (e.g., Haugb{\o}lle et al. 2018, Bate 2019). In these models, the instantaneous CMF and the final stellar IMF are not one-to-one either.

~~~~~Importantly, the CMF has been barely measured across different environments, especially outside the Solar neighborhood. The ALMA-IMF project (PI. F. Motte), now underway, is an attempt to measure the CMF in fifteen star forming regions across the Galaxy. We also note that many current dust measurements of the CMF are somewhat uncertain, due to intrinsic challenges, such as temperature determinations and whether each core will form one star or more (for more details see Offner et al. 2014). The difficulties in assignment of emission to a single object using automated routines, and the potential bias in the resulting CMFs are discussed in Pineda et al. (2009).

\section{Does Environment Sculpt the IMF?}

The question of how environment shapes the IMF can be reduced to two principal variables: metallicity and stellar feedback.

\subsection{Metallicity Effects}

Metallicity sets the opacity of the cloud, which governs the minimum mass of a gravitationally bound core, the cooling rate of cores, and the maximum possible stellar mass (Eddington luminosity). Therefore, we expect metallicity to play a pivotal role in shaping the IMF. Indeed, recent observations of early-type galaxies find that their local IMFs become increasingly bottom-heavy (i.g. more lower mass stars) in those galaxies that are metal rich (Martin-Navarro et al. 2015). 

\subsection{Feedback Effects}

According to simulations by Krumholz et al.\ (2016), radiative heating is the main driver of the characteristic IMF mass. These simulations show that when radiative heating increases, the efficiency of fragmentation is reduced,  leading to a top-heavy IMF. Conversely, Conroy \& van Dokkum (2012) suggest that a pivotal role is played by radiative ambient pressure, which is responsible for giving rise to bottom-heavy IMFs (at increasing pressure) as observed in elliptical galaxies with a history of starburst-generating mergers. An additional effect is represented by kinetic feedback. Stellar winds, protostellar outflows/jets, and ionization all likely affect the efficiency of star formation (e.g. Li $\&$ Nikamura 2006). However, it is still matter of debate how and if they ultimately affect stellar masses. For instance, it is thought that outflows slow the star formation rate (e.g. Dale $\&$ Bonnell 2008; Wang et al.\ 2010), but it is unclear if this has any effect on the stellar mass distribution. Likewise for ionization, several studies (e.g. Dale $\&$ Bonnell 2012; Walch et al.\ 2013) have shown that ionizing radiation can provide both negative or positive feedback, in the sense of suppressing or triggering star formation, but none of these have been conclusive in demonstrating the impact on the IMF.

\section{Open Questions}

{\bf{\underline{Overarching Questions:}}}

\begin{enumerate}
\item{To what extent can we assume the IMF is universal?}
\item{Does the CMF map directly on to the IMF in all environments?}
\item{How does environment shape the CMF and IMF?}
\end{enumerate}

\noindent
{\bf{\underline{More Specific Questions:}}}\\

\begin{enumerate}
\item{What physical mechanism(s) suppresses the formation of brown dwarfs? Can this lead to a better understanding of the distinction between brown dwarfs and giant planets?}
\item{Hierarchical collapse models and many observations suggest that giant molecular cloud (GMC) complexes make stars over an extended time period. Can we observe time-evolution in the CMF?}
\item{Massive stars do not seem to obey the CMF--IMF mapping. The CMF appears lognormal, not a Salpeter power-law slope at high masses. Do star formation efficiencies change at higher masses or are cloud mergers required to form the most massive stars?}
\item{How do binary/multiple stellar systems arise from the CMF? What determines whether gravitationally-bound cores fragment further?}
\item{Can we reconcile observations of bottom-heavy IMFs in elliptical galaxies that were once starbursts with top-heavy IMFs in young massive clusters (YMCs)?   What are the implications for Pop III stars?}
\end{enumerate}

\section{Observational Goals and Recommendations}

To answer the questions above we outline the following observational goals and recommendations:

\begin{itemize}
    \item{{\bf{Observational Goal---IMF:}} 
    To achieve an accurate measurement of the IMF in diverse environments and explore potential variations with metallicity and feedback, we require observations of 
    numerous YMCs and associations distributed at increasing distances across the Galaxy and beyond, such as Taurus ($d=180$ pc); Orion (400 pc); M17 (1.6 kpc); W3/4/5 (2.0 kpc, outer Galaxy); NGC 7538 (2.8 kpc), NGC 3603 (7 kpc), and the Large and Small Magellanic Clouds (50--60 kpc).\\
    
    \hspace{0.5truecm}
    
    {\bf{Recommendation:}} While {\em Gaia} can provide information on distances and velocities for the stars in these star-forming complexes, 
    we need high spatial-resolution (${<}0.1"$), wide-field imaging and spectroscopy at visual and particularly NIR wavelengths to allow stellar age and mass determinations in both unobscured and obscured regions up to several degrees wide on the sky. This type of information can be obtained by the {\em Cosmological Advanced Survey Telescope for Optical and ultraviolet Research} ({\em CASTOR}, Cot\'e et al.\ 2012) and by {\em WFIRST}. These facilities, combined with LSST ($u,g,r,i,z, Y$) and Euclid ($R,I,Z, Y, J, H$), will be ideal for studies of the IMF, thanks to their wavelength coverage (0.15 - 0.4 $\mu$m for {\em CASTOR}/Visible Imager and Spectrometer and 0.4 - 2 $\mu$m for {\em WFIRST}/WFI) and large FOV (0.67 deg$^{2}$ for {\em CASTOR}/Visible Imager, and 0.25 deg$^{2}$ for {\em WFIRST}/WFI).}
    
    \item{{\bf{Observational Goal---CMF:}} Along the same lines as for the IMF, we advocate for surveys of Galactic and extra-galactic GMC complexes (as described above) to investigate potential variations of the CMF with environment.\\ 
    
    \hspace{0.5truecm}
   
    {\bf{Recommendation:}}  Interferometric observations (ALMA, EVLA, SMA) will provide high-resolution observations of targeted regions. However, the {\em Origins Space Telescope (OST)} will be uniquely capable of performing statistical measurements of the CMF and protostellar luminosity functions in distant/obscured Galactic regions, including starburst-like environments. What makes \OST\ ideal for this task is its unique imaging and mapping capabilities of the far-IR cold dust emission peak in dense, prestellar clumps and cores. This can be achieved through the combination of (1) a large FOV for efficient scanning of extended regions on the sky; (2) sufficiently high angular resolution to resolve a 0.1 pc cores at a distance of a few kpc, and (3) sufficiently high sensitivity to detect 0.1~M$_{\odot}$ cores at 2--3 kpc. The 5.9-m \OST\ mirror allows achieving a resolution of ${\sim}6''$ at 50~$\mu$m, which is comparable to the angular resolution at shorter wavelengths of {\em Spitzer}/IRAC ($2''$), and {\em Spitzer}/MIPS ($6''$). The Far-Infrared Imager and Polarimeter instrument, FIP, can map 1~deg$^{2}$ of the sky in 100~hrs while achieving a 5-$\sigma$ sensitivity of ${\sim}1~\mu$Jy. We note that the baseline concept for the \OST/FIP instrument has two bands---50 and 250~$\mu$m---but an optional upscope would add the 100 and 500~$\mu$m channels. We recommend the inclusion of these additional bands that would better constrain the peak of the cold dust emission. 
    }
\end{itemize}

\noindent
Importantly, currently existing (e.g. {\em HST}, the Magellan telescope, etc.) or planned facilities (e.g., the {\em James Webb Space Telescope}) will be able to carry out imaging and multi-object spectroscopy of targeted Galactic YMCs. While such observations will be useful for IMF studies, the reach of these measurements will be limited by the small FOVs, which do not allow efficient mapping of large (i.e. of the order of deg$^{2}$) star-forming complexes across the Galaxy.

\pagebreak
\textbf{References}\\

\small{
Andr\'e, P., et al.\, 2010, A$\&$A, 518, 102\\
Bastian, N., et al.\, 2010, ARA$\&$A, 48, 339\\
Bate, M. 2019, MNRAS, 484, 2341\\
Cappellari, M., et al.\, 2012, Nature, 484, 485\\
Chabrier, G., et al.\, 2003, PASP, 115, 763\\
Chabrier, G., 2005, {\em{The Initial Mass Function 50 Years Later}}, vol. 327 of Astrophysics and Space Science Library, ed. by Corbelli, Palla $\&$ Zinnecker, pp. 41-50, Springer, Dordrecht\\
Chomiuk, L. $\&$ Povich, M., 2011, ApJ,142, 197\\
Conroy, C. $\&$ van Dokkum, P. G., 2012, ApJ, 760, 71\\
Dale, J. E. $\&$ Bonnell, I. A., MNRAS, 2008, 391, 2\\
Dale J. E. $\&$ Bonnell, I. A., MNRAS, 2012, 422, 1352\\
Dave', R., 2008, MNRAS, 385, 147\\
Elmegreen, B. G., 2011, ApJ, 564, 773\\
Ferreras, I., et al.\, 2013, MNRAS, 429, 15\\
Haugb{\o}lle, T., Padoan, P., Nordlund, \AA. 2018, ApJ, 854, 35\\
Hennebelle, P. $\&$ Chabrier, G., 2008, ApJ, 684, 395\\
Hosek, M., et al.\, 2019, ApJ, 870, 44\\
Hussman, B., et al.\, 2012, A$\&$A, 540, 57\\
Inutsuka, S.-I. 2001, ApJ, 559, L149\\
Johnstone, D., et al.\, 2000, ApJ, 545, 327\\
Konyves, V., et al.\, 2015, A$\&$A, 584, 81\\
Kroupa, P., 2001, MNRAS, 322, 231\\
Kroupa, P., 2002, Science, 295, 82\\
Kroupa, P., 2013, {\em{The Stellar and Sub-Stellar Initial Mass Function of Simple and Composite Populations}}, p. 115\\
Krumholz, M., et al.\, 2016, MNRAS, 458, 1671\\
Li, Z.-Y. $\&$ Nakamura, F., 2006, ApJ Letters, 640, 187\\
Lim, B., et al.\, 2013, AJ, 145, 2, 46\\
Lu, J. R., et al.\, 2013, ApJ, 764, 155\\
Martin-Navarro, I., et al.\, 2015, ApJL, 806, 31\\
Motte, F., et al.\, 1998, A$\&$A, 336, 150\\
Motte, F., et al., 2018, Nature Astronomy, 2, 478\\
Offner, S., et al.\, 2014, {\em{The Origin and Universality of the Initial Mass Function}}, Protostars and Planets IV, ed. Beuther, Klessen, Dullemond, Henning, University of Arizona Press, 914 pp., p. 53-75\\
Olmi, L., et al.\, 2018, MNRAS, 480, 1831\\
Pineda, J. E., Rosolowsky, E., Goodman, A. A., 2009, ApJ, 699, 134 \\
Shadmehri, M $\&$ Elmegreen, B. G., 2011, MNRAS, 410, 788\\
Testi, L. $\&$ Sargent, A. I., 1998, ApJL, 508, 91\\
Treu, T., et al.\, 2010, ApJ, 709, 1195\\
Van Dokkum, P. G. $\&$ Conroy, C., 2010, nature, 468, 940\\
Walch, S., et al.\, 2013, mnras, 435, 917\\
Wang, P., et al.\, 2010, ApJ, 709, 27\\
}

\end{document}